%% file: SC_SSE.tex
\documentclass{PoS}

\usepackage[pdftex]{graphicx}
\usepackage[dvips]{epsfig}
\usepackage{bbm}
\usepackage{verbatim}
\usepackage{psfrag}
\usepackage{slashed}

\usepackage{amsmath}
\usepackage{amssymb}
\usepackage{marvosym}

\usepackage{caption}

\usepackage{arydshln}

\def \Nt {{N_{\tau}}}
\def \at {{a_{\tau}}}

\def \hmu {{\hat{\mu}}}

\newcommand{\beqn} {\begin{equation}}
\newcommand{\eqn} {\end{equation}}

\def \beq{\begin{equation}}
\def \eeq{\end{equation}}
\def \bea{\begin{eqnarray}}
\def \eea{\end{eqnarray}}

\def \Tr {{\rm Tr}}

\def \bet0{\beta_0}
\def \bet1{\beta_1}
\def \simgt{\,\rlap{\lower 7.5 pt\hbox{$\mathchar \sim$}}\raise 3 pt \hbox{$>$}\,}
\def \simlt{\,\rlap{\lower 7.5 pt\hbox{$\mathchar \sim$}}\raise 3 pt \hbox{$<$}\,}

\def\lsim{\raise0.3ex\hbox{$<$\kern-0.75em\raise-1.1ex\hbox{$\sim$}}}

\input{newcommand.tex}

\title{New algorithms and new results\\ for strong coupling LQCD}

\ShortTitle{New algorithms and new results for SC-LQCD}

\author{\speaker{Wolfgang Unger}\\
   Institut f\"ur Theoretische Physik, Goethe-Universit\"at Frankfurt, 60438 Frankfurt am Main, Germany\\
        E-mail: \email{unger@th.physik.uni-frankfurt.de}
}

\author{Philippe de Forcrand\\
        Institut f\"ur theoretische Physik, ETH Z\"urich, CH-8093, Switzerland\\
        CERN, Physics Department, TH Unit, CH-1211 Geneva 23, Switzerland\\
        E-mail: \email{forcrand@phys.ethz.ch}
}

\abstract{
\vspace*{-10.5cm}
\begin{flushright}
\texttt{\footnotesize CERN-PH-TH/2012-326}\\
\end{flushright}
\vspace*{9.5cm}
We present and compare new types of algorithms for lattice QCD with staggered fermions in the limit of infinite gauge coupling.
These algorithms are formulated on a discrete spatial lattice but with continuous Euclidean time.
They make use of the exact Hamiltonian, with the inverse temperature beta as the only input parameter.
This formulation turns out to be analogous to that of a quantum spin system. The sign problem is completely absent, at zero and non-zero baryon density.
We compare the performance of a continuous-time worm algorithm and of a Stochastic Series Expansion algorithm (SSE),
which operates on equivalence classes of time-ordered interactions.
Finally, we apply the SSE algorithm to a first exploratory study of two-flavor strong coupling lattice QCD,
which is manageable in the Hamiltonian formulation because the sign problem can be controlled.
}

\FullConference{The 30 International Symposium on Lattice Field Theory - Lattice 2012,\\
		June 24-29, 2012\\
		Cairns, Australia}

\begin{document}

\vspace{-1mm}
\section{Introduction}
\vspace{-2mm}

Lattice QCD with staggered fermions in the strong coupling limit (SC-LQCD) is a useful effective model of QCD, as it shares important features of QCD such as confinement and spontaneous chiral symmetry breaking and 
its restoration at a transition temperature $T_c$.
One can study the nuclear potential as well as the phase diagram at non-zero baryon chemical potential $\mu_B$ \cite{Forcrand2010}. 
These topics can not be properly addressed with conventional, determinant-based lattice QCD using HMC algorithms, due to the notorious sign problem:
all methods available today are limited to $\mu_B/T \lesssim 1$ \cite{Forcrand2009}. 
In contrast, SC-LQCD can be reformulated as a monomer-dimer
system \cite{Rossi1984}. There, the sign problem can be made mild due to a resummation of baryonic and mesonic degrees of freedom \cite{Karsch1989}.
Due to algorithmic developments over the last decade, in particular due to the application of the Worm algorithm to the monomer-dimer partition function \cite{Adams2003}, 
SC-LQCD - which has been studied via mean field theory \cite{Meanfield} and with Metropolis algorithms \cite{Rossi1984, Karsch1989} since the 1980s -  has
experienced a revival, as simulations at finite baryon density could be performed with modest computational demands. 
Moreover, the chiral limit can be studied very economically - simulations are faster than with a finite quark mass.
However, limitations remain. In particular, only the 1-flavor (4 tastes) theory has been considered so far in the dimer-formulation. 
The physically more interesting case of 2 flavors could not be addressed yet 
due to a severe sign problem in the mesonic sector \cite{FrommThesis}.
Here, we propose a Hamiltonian formulation of strong coupling lattice QCD based on the Euclidean continuous time limit, where further simplifications occur.
In particular, we show that this formulation is a generalization of Hamiltonians for spin systems. It can in principle be extended to arbitrary $\Nf$. 
In this paper, we illustrate the formalism and give first Monte Carlo results obtained via Stochastic Series Expansion for $\Nf=2$ and U(2) gauge group. 

\section{The continuous Euclidean time approach}

\newcommand{\tbfb}[1]{{\emph{#1}}}

In SC-LQCD, the gauge coupling is sent to infinity and hence the coefficient $\beta=2\Nc/g^2$ of the plaquette term representing the Yang Mills part $F_{\mu\nu} F_{\mu\nu}$ of the action is zero.
The lattice becomes maximally coarse, and no continuum limit can be considered.
But the gauge fields in the covariant derivative can be integrated out analytically because the integration factorizes. After the Grassmann integration over the fermions, one obtains the 
SC-LQCD partition function \cite{Rossi1984} in the dimer representation, which is an exact rewriting of the 1-flavor staggered fermion action on a $d+1$ dimensional lattice $N_\sigma^d\times N_\tau$:
\begin{align}
S[U,\chi,\bar{\chi}]=am_q\sum_x\bar{\chi}(x) \chi(x) 
&+ \frac{\gamma}{2}\sum_x \eta_0(x) \left[ \bar{\chi}(x) e^{\at\mu} U_0(x)\chi(x+\hat{0})- \bar{\chi}(x+\hat{0}) e^{-\at\mu} U_0^\dagger(x)\chi(x)\right]\nn
&+ \frac{1}{2}\sum_x  \sum_i^d \eta_i(x) \left[ \bar{\chi}(x) U_i(x)\chi(x+\hat{i})- \bar{\chi}(x+\hat{i}) U_i^\dagger(x)\chi(x)\right]
\label{SCQCDPF}\\
\longrightarrow \qquad Z(m_q,\mu_q)=& \sum_{\{k,n,\ell\}}'\prod_{b=(x,\hat{\mu})}\frac{(\Nc-k_b)!}{\Nc!k_b!}
\gamma^{2k_b\delta_{\hat{0}\hat{\mu}}}\prod_{x}\frac{\Nc!}{n_x!}(2am_q)^{n_x} \prod_\ell w(\ell),\nonumber \\
&\qquad w(\ell)=\sigma(\ell) \gamma^{\Nc \sum_x |\ell_0(x)|}\exp(\Nc \Nt r_\ell a_\tau \mu),
\label{SCPFF}
\end{align}
with $m_q$ the quark mass and $\mu=\frac{1}{\Nc}\mu_B$ the quark chemical potential, $\sigma(\ell)=\pm 1$ a geometry dependent sign and $r(\ell)$ the winding number of baryon loop $\ell$.
The sum $\sum\nolimits'$ is over admissible configurations, namely those which fulfill the Grassmann constraint 
\begin{equation}
 n_x+\sum_{\hat{\mu}=\pm\hat{0},\ldots \pm \hat{d}} \left(k_{\hat{\mu}}(x) +\frac{\Nc}{2} |\ell_\mu(x)|\right) =\Nc \quad 
\forall x\in V.
\label{GC}
\end{equation}
Since color degrees of freedom have been integrated out, configurations are defined in terms of mesons
- represented by the monomers $n_x\in\{0,\ldots \Nc\}$ and dimers $k_\mu(x)\in \{0,\ldots \Nc\}$ (non-oriented meson hoppings) 
- and baryons - represented by self-avoiding closed loops constituted by $\ell_\mu(x)\in\{-1,0,+1\}$.\footnote{Note that U(3) describes a purely mesonic system, while SU(3) contains baryons.}
Here, we consider the chiral limit, $m_q=0$ where monomers are absent: $n_x=0$.

In Eq.~(\ref{SCQCDPF}) we have introduced an anisotropy $\gamma$ in the Dirac couplings. This complication is necessary because the chiral restoration temperature
is given by roughly $a T\simeq1.5$, and on an isotropic lattice with $aT=1/\Nt$ we could not
reach sufficiently high temperatures. Furthermore, varying $\gamma$ is the only way to vary the temperature continuously.
The temperature is thus
 $aT=f(\gamma)/ \Nt$ with $f(\gamma)=a/\at$.
However, the functional dependence $f(\gamma)$ of the ratio of the spatial and temporal lattice spacings on $\gamma$ is not known.
Naive inspection of the derivatives in Eq.~(\ref{SCQCDPF}) would indicate $f(\gamma)=\gamma$, but this only holds at weak coupling.
In contrast, the mean field approximation of SC-QCD based on a $1/d$-expansion \cite{Bilic1992} 
suggests that $aT_c=\gamma_c^2\Nt$ is the sensible, $\Nt$-independent identification in leading order in $1/d$. We have emphasized elsewhere \cite{Unger2011} 
by analytic arguments and numerical investigation that this identification is the only suitable one 
which renders observables like the chiral susceptibility and the specific heat finite in the limit $\Nt, \gamma \rightarrow \infty$.
However, this limit (keeping $\gamma_c^2\Nt$ fixed) is approached with significant, sometimes non-monotonic $1/\Nt$ corrections.
To circumvent such extrapolation problems, we consider the \emph{continuous Euclidean time ({\rm CT}) limit:}
$\Nt\rightarrow \infty$, $\gamma \rightarrow \infty$, with $\gamma^2/\Nt\equiv aT$ fixed.
Hence we are left with only one parameter $\beta\equiv\Nt/\gamma^2$ to set the thermal properties, and all discretization errors introduced by a finite $\Nt$ are removed. 
Moreover, in the baryonic sector the partition function simplifies greatly: baryons become static in the CT limit, hence the sign problem is completely absent.
Additionally, multiple spatial dimers $k_i(x)>1$ become completely suppressed (see \cite{Unger2011}) and one can derive the CT partition function:
\begin{equation}
Z_{\rm CT}(\beta,\mu)=\sum_{k\in 2\mathbb{N}}(\beta/2)^{k}\sum_{\mathcal{G}'\in \Gamma_k} e^{3\mu\beta B} \hat{v}_L^{N_L}\hat{v}_T^{N_T}
\quad\text{with}\quad
k=\sum_{b=(x,\hat{i})} k_b=\frac{N_L+N_T}{2}, \quad N_{L/T}=\sum_x n_{L/T}(x)
\label{PARFCT}
\end{equation}
where $B$ is the baryon number, and $\Gamma_k$ is the set of equivalence classes $\mathcal{G}'$ of graphs containing a total number $k$ of spatial hoppings, equivalent up to time shifts of the vertices.
The vertex weights $v_L=1$ and $v_T=2/\sqrt{3}$ label L- and T-types of vertices as illustrated Fig.~1 (left).
An important property of the partition function Eq.~(\ref{PARFCT}) is that spatial dimers are distributed uniformly in time.
The lengths $\Delta\beta$ of ``dashed'' or ``solid'' time intervals (see Fig.~1 left) are then, according to a Poisson process, exponentially distributed:
$P(\Delta\beta)\propto \exp(-\lambda \Delta\beta)$, $\Delta\beta \in [0,\beta]$ with $\lambda$ the ``decay constant'' 
for spatial dimer emission  $\lambda=(2d-\sum_{\mu} n_B(x\pm\hmu))/4$. This is the basis for the continuous time Worm algorithm presented in \cite{Unger2011}.

\section{The 1-flavor Hamiltonian}

The Hamiltonian formulation can be obtained from Eq.~(\ref{PARFCT}) by realizing that the degrees of freedom can be mapped on a spin system. We can restrict the discussion to the mesonic sector
$U(\Nc)$, since baryons are static for $\Nc>2$.
Notice that, except for spatial hoppings, meson lines are time-like and form
dimer chains alternating between $k_0^{\rm even}(\vec{x}) \in \{0,\ldots \Nc\}$ and $k_0^{\rm odd}(\vec{x})$ with $k_0^{\rm odd}(\vec{x})=(\Nc-k_0^{\rm even}(\vec{x}))$
on even and odd time-slices.
One can then introduce the observable
\begin{equation}
S^z(\vec{x},t) =\frac{(-1)^{x+y+z+t}}{2}(2 k_0(\vec{x},t)-\Nc) \in \{-\Nc/2, \ldots \Nc/2\}
\end{equation}
which is constant on static lines. The ``spin'' $S_z$ simply counts the number of time-like meson hoppings, and is in no way related to the spin of the quarks.
Spatial dimers can then be oriented consistently, such that for each spatial dimer between a pair of neighboring sites 
$\expval{\vec{x},\vec{y}}$, one unit of spin $\Delta S^z=\pm 1$ is transferred from site $\vec{x}$ to site $\vec{y}$. 
Hence the total spin $S^z\equiv S^z(t) =\sum_{\vec{x}\in V}S^z(\vec{x},t)$ is globally conserved.

\begin{minipage}{0.32\textwidth}
\includegraphics[width=\textwidth]{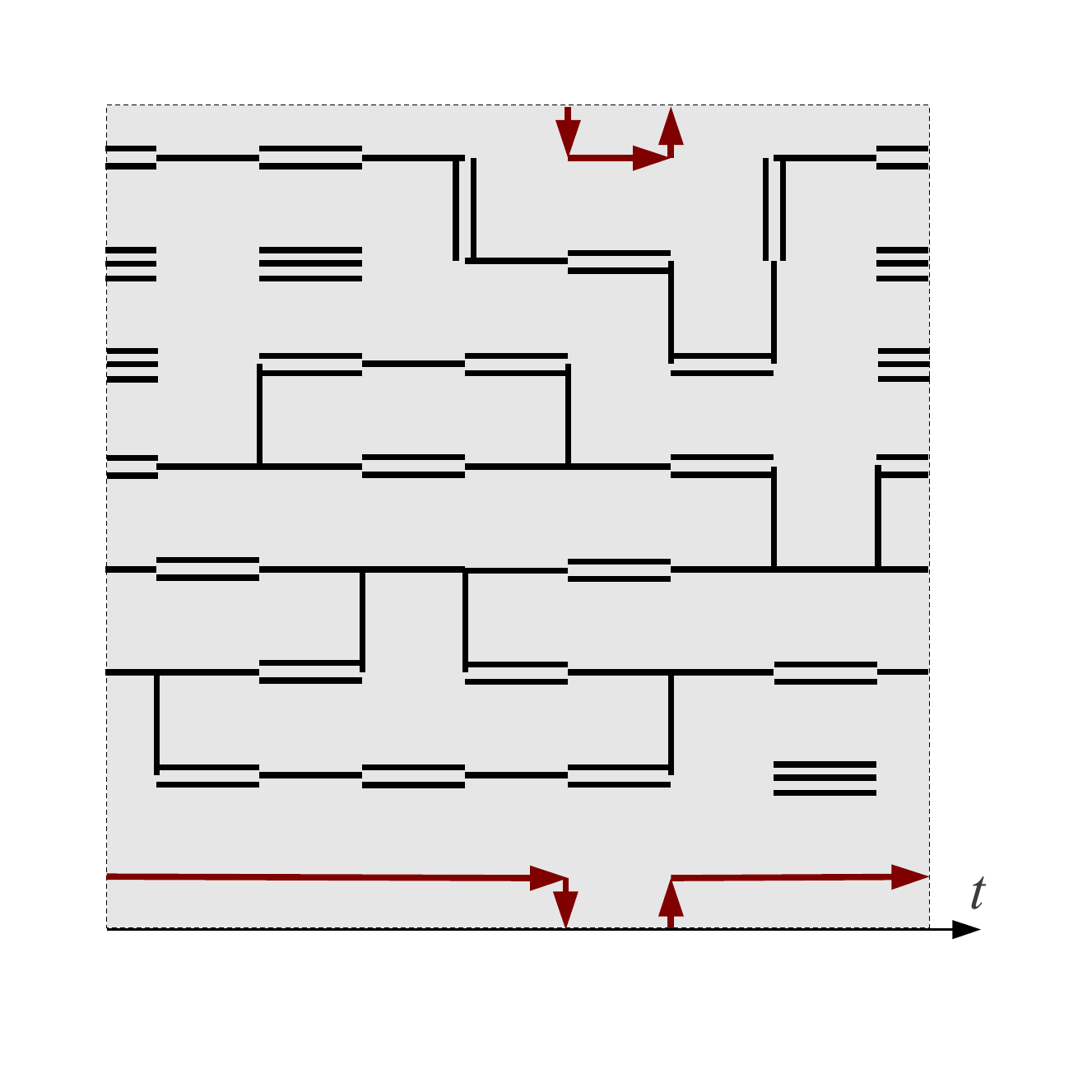}\\
\vspace{-13mm}\\
\includegraphics[width=\textwidth]{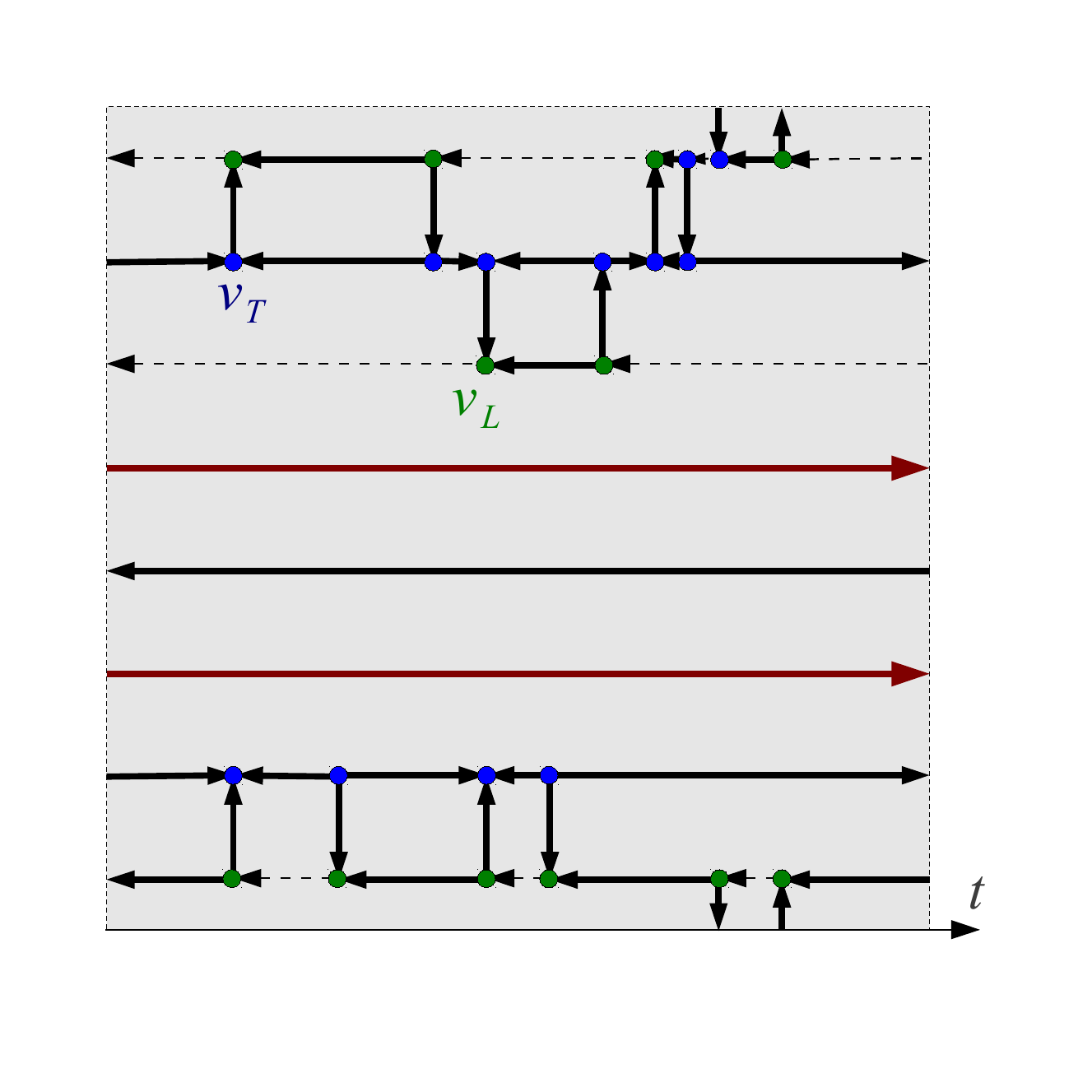}
\end{minipage}
\begin{minipage}{0.67\textwidth}
\vspace{-7mm}
\centerline{\includegraphics[width=0.8\textwidth]{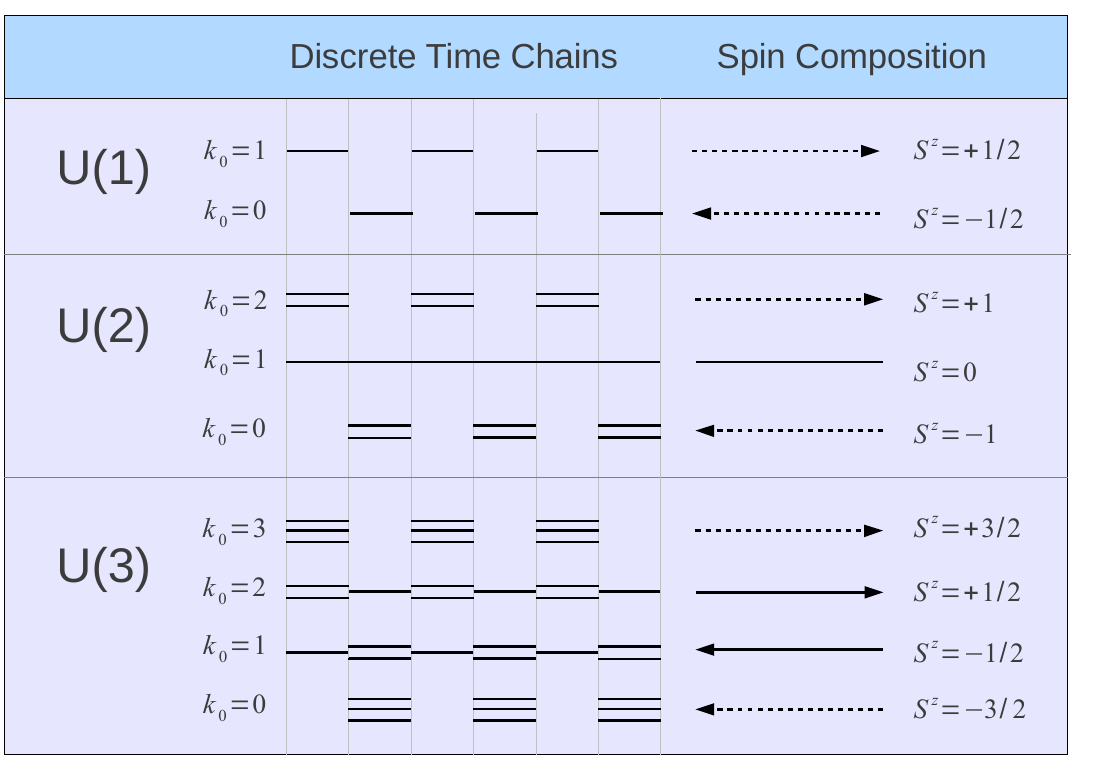}}
\captionof{figure}{Left: Typical 2-dimensional configurations in discrete (top) and continuous time (bottom) at the same temperature,
where multiple dimers are absent and baryons (red) become static. 
L- and T-vertices have different weights.
Meson lines can be oriented in a consistent way. Right: Mapping of static mesons (time-like chains) to a ``spin''. }
\end{minipage}

\newcommand{\diag}[1]{\rm diag\left(#1\right)}

In the case of gauge group U(1) (lattice QED in the strong coupling limit), the Hamiltonian is identical to that of the XY Model in zero field: $\hat{H}=\sum_{\expval{\vec{x},\vec{y}}} \sigma^+ \sigma^-$, 
with $\sigma^\pm=\sigma_1 \pm i\sigma_2$ the spin raising/lowering operators constructed from the Pauli matrices.
The generalization to $U(\Nc)$ reads
\begin{equation}
Z(\beta)=\Tr\left[e^{-\beta \hat{H}}\right], \qquad 
\hat{H}=-\frac{1}{2}\sum_{\expval{\vec{x},\vec{y}}}\left(J^+_{\vec{x}} J^-_{\vec{y}} + h.c.\right)
\qquad
J^{+}={\scriptsize \left(
\begin{array}{llll}
0 &  &  &  \\
v_1 & 0 &  & \\
 & \ddots & \ddots \\
 & & v_{\Nc} & 0\\
\end{array}
\right)},\quad v_k=\sqrt{\frac{k(1+\Nc-k)}{\Nc}}
\label{Hamiltonian}
\end{equation}
with $J^{-}=(J^+)^T$ which are spin lowering/raising operators. The off-diagonal matrix elements $v_k$ are generalized vertex weights. 
For $\Nc=3$, we can identify $v_L\equiv v_1=v_3=1,\; v_T\equiv v_2=2/\sqrt{3}$.
Note that the operators $J^\pm$ reflect the existence of a lowest and highest weight, $J^-|-\Nc/2\rangle=0$, $J^+|\Nc/2\rangle=0$,
and fulfill as well the commutation relation $\frac{\Nc}{2}[J^+,J^-]\equiv J^z=\diag{-\Nc/2,\ldots \Nc/2}$ with $J^z|S^z\rangle = S^z|S^z\rangle$. 
This justifies the characterization of static lines in terms of  a ``spin'' quantum number.

\section{Stochastic Series Expansion}

The partition function Eq.~(\ref{PARFCT}), which is a sum of weighted diagrams in a  perturbative series in $\beta$, 
can be sampled via diagrammatic Monte Carlo techniques such as the continuous time worm algorithm \cite{Beard1996} or the
\tbfb{Stochastic Series Expansion} (SSE) \cite{Sandvik1999}. Here, we restrict to SSE: as we will see, it can be easily generalized to $\Nf>1$ once we have constructed the corresponding Hamiltonian.
SSE is based on a rewriting of the partition function by inserting identity and diagonal matrix elements:
\begin{equation}
Z(\beta)=\Tr\left[e^{-\beta \hat{H}}\right]=\sum_{\chi}\sum_{\{S_L\}}\frac{\beta^\kappa(L-\kappa)!}{L!}\expval{\chi \left|\prod_{i=1}^L \hat{H}_{a_i,b_i} \right|\chi},\quad\hat{H}_{1,b}=c\Id,\quad \hat{H}_{2,b}=\frac{1}{2}\left(J^+_{\vec{x}} J^-_{\vec{y}} +h.c.\right)
\end{equation}
where $\chi$ is a state vector and $S_L$ is a time-ordered sequence of indices: $S_L=\{[a_1,b_1],[a_2,b_2], \ldots [a_L,b_L]\}$ characterizing - together with an initial state $\chi$ - a graph in $Z(\beta)$.
$L$ is the number of operators and $\kappa<L$ the order in the expansion in $\beta$.  
The indices $a_i=0$ correspond to the identity, $a_i=1$ to the diagonal matrix element $c\Id$, where $c$ can be adjusted in order to simplify the algorithm, and $a_i=2$ to non-diagonal matrix elements. The index
$b_i=\expval{\vec{x},\vec{y}} \in Vd$ denotes a bond on the lattice, and for $a_i=0$, $b_i=0$ denotes a dummy bond.
For any finite volume and given temperature, only a finite number of orders in $\beta$ contribute. 
$L$ can be set to be larger than this number, making SSE approximation-free.	
The algorithm consists of two kinds of updates: (1) a Metropolis update changing the order in $\beta$:
\begin{equation}
P([0,0] \mapsto [1,b])=\frac{N_\sigma^3 d\beta \expval{\chi|\hat{H}_{1,b}|\chi}}{L-\kappa},\quad  
P([1,b] \mapsto [0,0])=\frac{L-\kappa+1}{N_\sigma^3 d\beta \expval{\chi|\hat{H}_{1,b}|\chi}},
\end{equation}
and (2) the operator loop update, visiting a set of bonds in succession, starting from an input leg and determining the output leg with heatbath probability $\propto \expval{\chi(x)\chi(y)|\hat{H}_{a_i b_i}|\chi'(x)\chi'(y)}$ .
In Fig.~\ref{SSE} we compare different observables obtained from SSE and/or CT-Worm. We want to stress that a new observable, the spin susceptibility 
$\chi_S=\frac{\beta}{V}\expval{(\sum_i S_i^z)^2}$,
is also sensitive to the chiral transition.
It measures the fluctuations in the number of time-like mesons, and is thus
analogous (but not equal) to the specific heat.

\begin{figure}[h!]
\centerline{\includegraphics[width=0.8\textwidth]{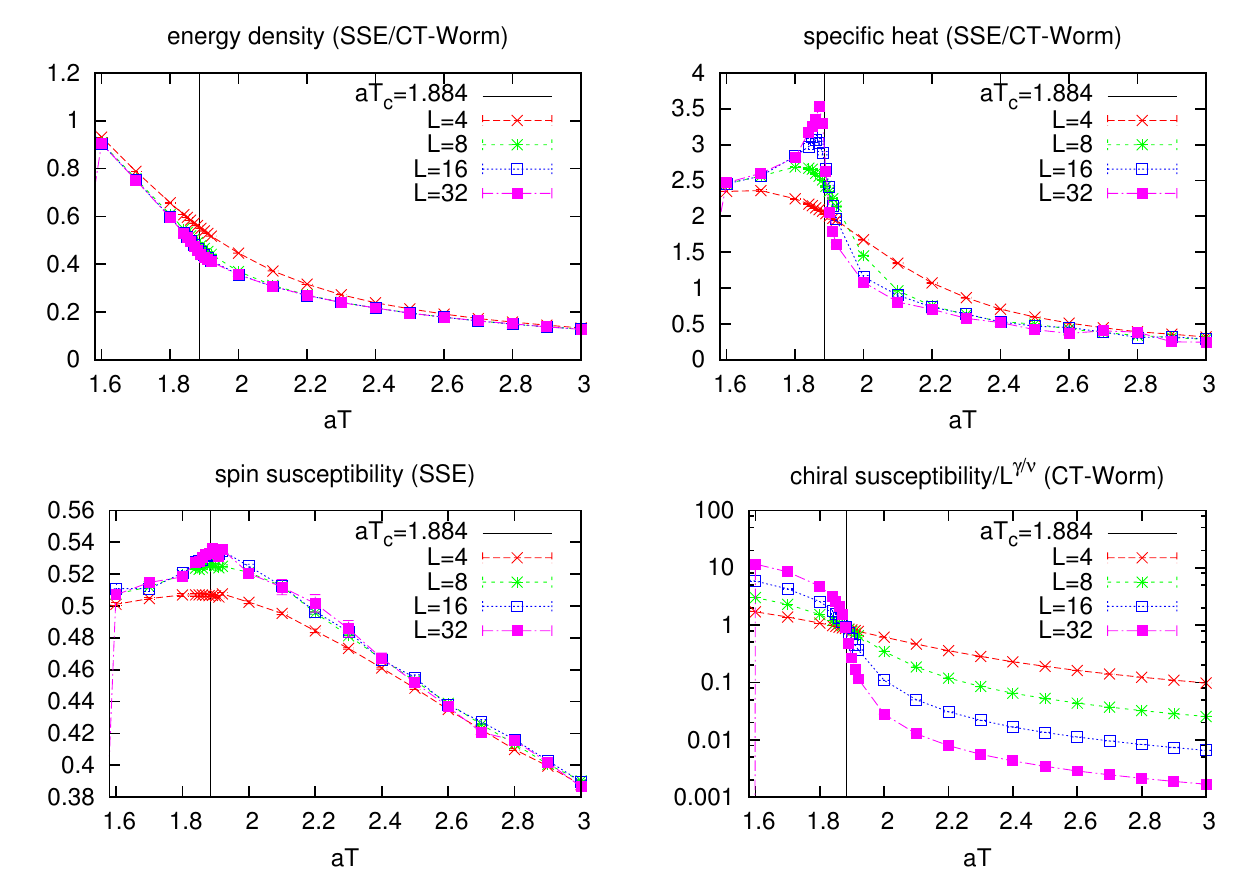}}
\caption{Observables in 1-flavor U(3): (top row) energy density and susceptibility measured with CT-worm and SSE agree; (bottom left) spin susceptibility measured with SSE; (bottom right) chiral susceptibility measured with CT-Worm.
\label{SSE}
}
\label{SSE}
\end{figure}

\section{The 2-flavor formulation and first results for U(2) gauge group}
\begin{figure}
\includegraphics[width=\textwidth]{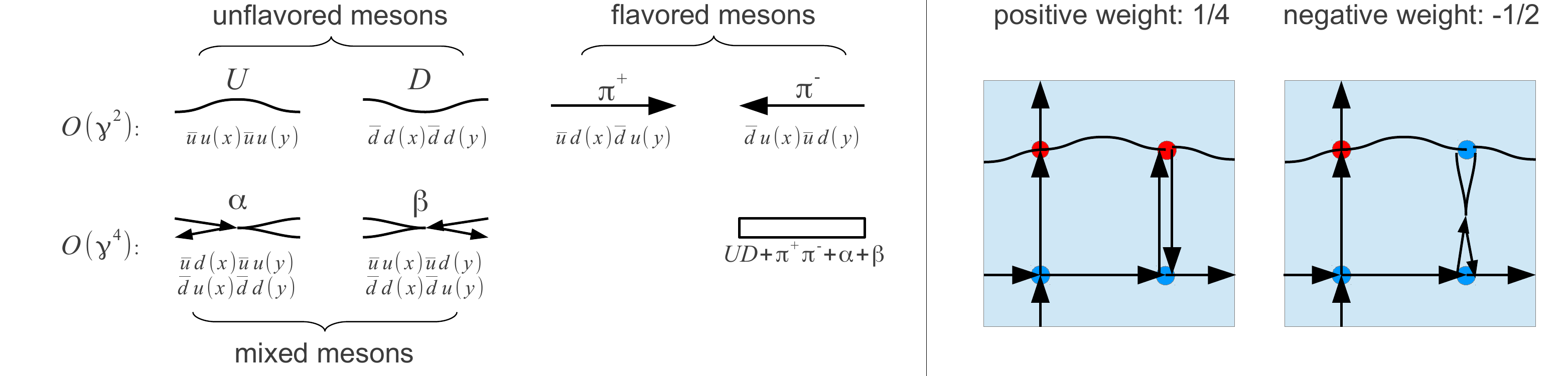}
\caption{Left: definition of unflavored, flavored, and mixed dimers. Right: 2x2 sample configurations with positive and negative weight.}
\label{Mixed}
\end{figure}

In the strong coupling limit, only one of the four tastes of staggered fermions remains light, since taste splitting is maximal. 
It was argued in \cite{Forcrand2010} that the nuclear interaction in 1-flavor SC-LQCD is due to entropic forces: the presence of static baryons
modifies the pion bath, with the modulation proportional to the $\rho$-propagator. 
However, with 1 flavor pion exchange cannot not occur, because mesons do not couple to baryons. Pion exchange can only be studied by going to $\Nf>1$.
But no 2-flavor formulation of staggered SC-QCD suitable for finite-density Monte Carlo exists, as already the mesonic sector has a severe sign problem \cite{FrommThesis}.
After reviewing the problems with the conventional dimer representation of SC-QCD, we derive a sign-problem-free Hamiltonian formulation in continuous Euclidean time.

The 1-link integrals which appear in the strong coupling limit can be expressed in terms of the gauge-invariant terms \cite{Eriksson1981}. 
 For SU(2) and U(2), the 1-link integals (valid for any $\Nf$) are 
\begin{equation}
 z_{SU(2)}(x,y)=\sum_{n=0}^{\infty} \frac{(X+\Delta)^n}{n!(n+1)!}, \qquad
 z_{U(2)}(x,y)=\sum_{i,j=0}^{\infty} \frac{X^i+Z^j}{(i+2j+1)!i!(j!)^2}
\label{OneLink}
\end{equation}
 with $X=tr(mm^\dagger)=U+D+\pi^+ + \pi^-$ the sum of $\Nf=2$ \tbfb{unflavored} (U=$\bar{u}u$ and D=$\bar{d}d$) and $\Nf(\Nf-1)=2$ \tbfb{flavored} mesons, 
$\Delta=\det(m)+\det(m^\dagger)$ the diquark+anti-diquark term, and $Z=\det(m m^\dagger)=X^2-UD-\pi^+\pi^- +\alpha +\beta$ a mesonic term.
Here, $\alpha=\bar{u}d_x\bar{d}u_x \bar{u}u_y \bar{d}d_y$ and $\beta=\bar{u}u_x \bar{d}d_x\bar{u}d_y\bar{d}u_y$ are the potentially problematic contributions,
which correspond to the mixing of UD and $\pi^+\pi^-$ dimer pairs (obtained via non-trivial Wick-contractions), as illustrated Fig.~\ref{Mixed}.
In particular, if a configuration contains an odd number of $\alpha$ or $\beta$ links, 
according to the Grassmann constraint (see \cite{FrommThesis}) the configuration has a negative sign.
The essential feature of the continuous Euclidean time formulation is that multiple spatial dimers, and hence also $\alpha$ and $\beta$ spatial dimers, are suppressed.
They can only enter in static lines, where they can be resummed with other static lines so that the sign problem disappears completely.
Combining time-like dimers of alternating orders is analogous to the procedure discussed in the $\Nf=1$ case \cite{Unger2011}: Chains of alternating orders 
$\Ord{\gamma^{2k}}\times \Ord{\gamma^{2\Nc\Nf-2k}}$ 
are resummed in a way consistent with the constraint Eq.~(\ref{GC}). 
This gives rise to new conserved quantum numbers: the spin is now composed of $\Nf$ separate spins
$S^z= \sum_{f=1}^{\Nf} S^z_f \in \{-\Nf \Nc/2, \ldots \Nf \Nc/2 \}$,
with $S^z_f\in \{-\Nc/2, \ldots \Nc/2\}$ each being conserved, and we also get $\Nf(\Nf-1)/2$ charges $Q_i\in \{-\Nc,\ldots,+\Nc \}$.
In the specific case of two flavors ($f=U,D$ and $Q_1=Q_\pi^\pm$), the spins $S_U, S_D$ can be replaced by $S^z$ and $Q_I=S^z_U-S^z_D \in \{-\Nc,\ldots,+\Nc\}$ which may be viewed as isospin. Here,
we find in total 19 types of states, as illustrated in Fig. \ref{classification}.\footnote{
For SU(2) we also have 3 kinds of diquarks, $uu$, $dd$ and $ud$, 
which are not suppressed in the CT-limit. To avoid this complication,
for $\Nc=2$ we restrict to U(2). For $\Nc=3$ this restriction is not necessary.
}
The state vector and the transition rules at spatial dimers are given by
\begin{equation}
\chi=\bigotimes_{\vec{x}\in V} \left|\{S_f^z(\vec{x})\}_{f=1,\ldots\Nf},\{Q_i(\vec{x})\}_{i=1,\ldots \Nf(\Nf-1)/2}\right\rangle,\qquad
|\Delta S^z|=1 \quad \text{and}\quad
\sum_i|\Delta Q_i|\in \{0,1\}. 
\end{equation}

\newcommand{\pp}[1]{\hspace{-1mm} #1 \hspace{-1mm}}

\newcommand{\ppa}{\hspace{-1mm} {U}}
\newcommand{\ppb}{\hspace{-1mm} {D}}
\newcommand{\ppc}{\hspace{-1mm} {\pi^+}}
\newcommand{\ppd}{\hspace{-1mm} {\pi^-}}

\newcommand{\qqa}{\hspace{-1mm} {U}}
\newcommand{\qqb}{\hspace{-1mm} {D}}
\newcommand{\qqc}{\hspace{-1mm} {\pi^+}}
\newcommand{\qqd}{\hspace{-1mm} {\pi^-}}

\newcommand{\pppa}{\hspace{-1mm} {\hat{U}}}
\newcommand{\pppb}{\hspace{-1mm} {\hat{D}}}
\newcommand{\pppc}{\hspace{-1mm} {\hat{\pi}^+}}
\newcommand{\pppd}{\hspace{-1mm} {\hat{\pi}^-}}

\newcommand{\qqqa}{\hspace{-1mm} {\hat{U}}}
\newcommand{\qqqb}{\hspace{-1mm} {\hat{D}}}
\newcommand{\qqqc}{\hspace{-1mm} {\hat{\pi}^+}}
\newcommand{\qqqd}{\hspace{-1mm} {\hat{\pi}^-}}

The Hamiltonian for $\Nf=2$ is now a sum of four contributions, implementing these transition rules:
\begin{equation}
\hat{H}=\frac{1}{2}\sum_{\langle \vec{x},\vec{y} \rangle}
\left( 
J^+_{U(\vec{x})}J^-_{U(\vec{y})}
+J^+_{D(\vec{x})}J^-_{D(\vec{y})}
+J^+_{\pi^+(\vec{x})}J^-_{\pi^+(\vec{y})}
+J^+_{\pi^-(\vec{x})}J^-_{\pi^-(\vec{y})} + h.c.
\right)
\end{equation}

\begin{minipage}{0.66\textwidth}
\includegraphics[width=\textwidth]{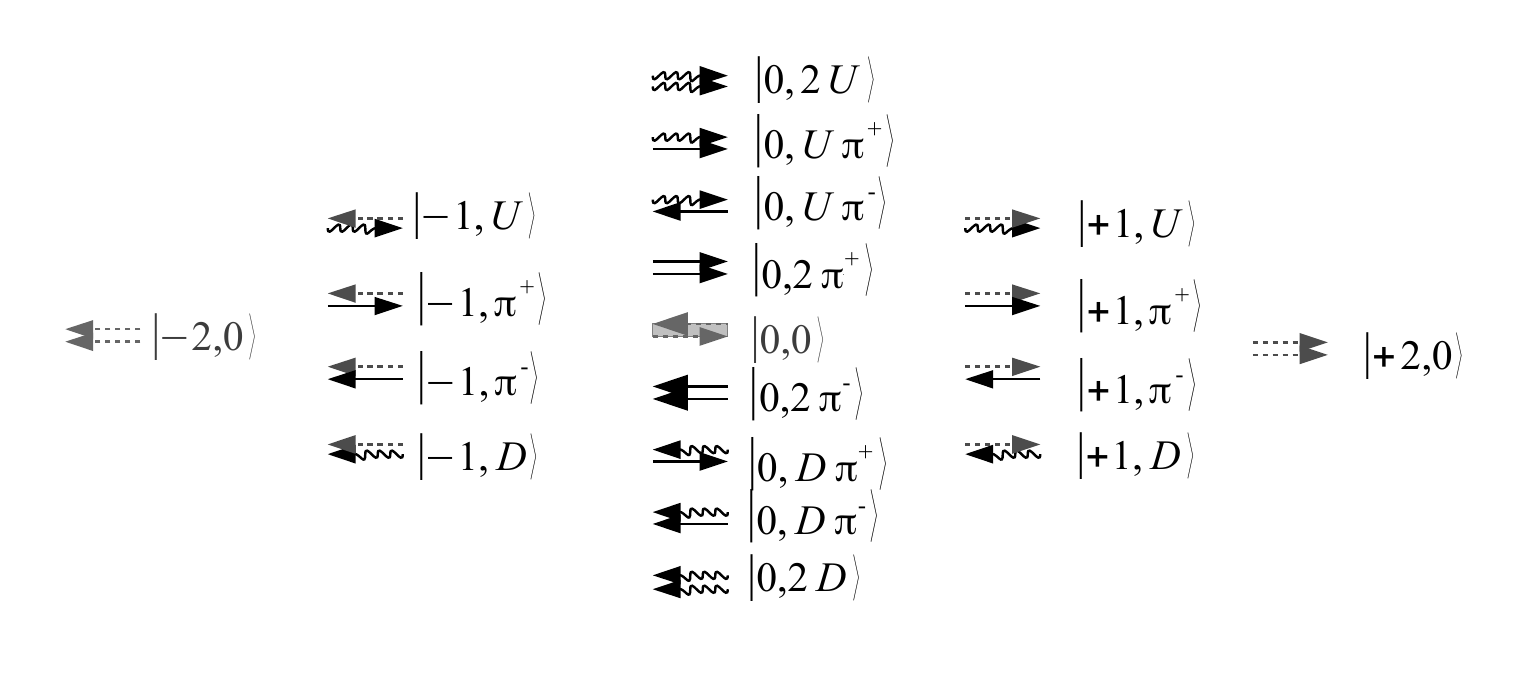}
\end{minipage}\quad
\begin{minipage}{0.3\textwidth}
\captionof{figure}{Classification of all static lines for U(2) based on ``spin'' quantum number and particle content. Each static line consists of two arrows, dashed arrows denoting spin, curly arrows denoting flavor-neutral mesons content, and solid arrows denoting charged mesons.}
\label{classification}
\end{minipage}

The absorption $J^+_{\pi_i}$, and emission $J^-_{\pi_i}$ operators can be represented as $19 \times 19$ lower left/upper right triangular matrices, where the entries are again given by 
 vertex weights: $v_{\hat{\pi}_i}=\frac{1}{\sqrt{2}}$ if states with $S=0$ and $|Q|=1$ are connected to the vertex, $v_{\pi_i}=1$ otherwise. This Hamiltonian can be used in the SSE algorithm to obtain results on the 
chiral phase transition. Our preliminary $\Nf=2$ results are compared to mean field predictions in Tab.~\ref{MFMC}.

\vspace{3mm}
\begin{minipage}{0.36\textwidth}
\centerline{$\begin{array}{rlll}
\hline
 \Nc & &\Nf=1 & \Nf=2\\
 \hline
 1 & & 3/2\; [1.102(1)] & 5/5\; [0.77(1)]  \\
 2 & & 4/2\;[1.467(1)] & 6/5\, [1.04(1)]  \\
 3 & & 5/2\; [1.884(1)] &  7/5 \\
\hline
\end{array}$}
\vspace{3mm}
\end{minipage}\qquad
\begin{minipage}{0.58\textwidth}
\captionof{table}{Comparison of the critical temperature $aT_c$ between mean field results and Monte Carlo results [in brackets] for ${\rm U}(\Nc)$ gauge groups. 
The new results are in column $\Nf=2$.
The MC value for $\Nf=2$, U(3) has not been measured yet.}
\label{MFMC}
\end{minipage}


\section{Conclusion}

We have given a new, Hamiltonian formulation of strong coupling lattice QCD with staggered fermions in the chiral limit. It is based
on the insight that strong coupling lattice QCD in the continuous time limit is analogous to a spin system. A new observable, 
the spin susceptibility, turns out to be sensitive to the chiral transition. Also, the Hamiltonian description allows to apply quantum Monte Carlo methods. In \cite{Unger2011} we have studied 1-flavor thermodynamics via the continuous time Worm algorithm.
Here, we make the first step towards 2-flavor simulations, by making use of Stochastic Series Expansion, a diagrammatic Monte Carlo technique which we generalize to
the Hamiltonian in question. SSE has the advantage that more complicated Hamiltonians can be studied with ease. The drawback of SSE, in contrast to the continuous time Worm, is that we do not know of a way to obtain 
the two-point function for free. For the computation of the specific heat, the performance of both algorithms is quite similar.
We have provided first results on the U(2) transition with two flavors. The extension to SU(3) with finite baryon chemical potential is more involved: the number of static lines 
increases to 44 in the mesonic sector.

\vspace{-1mm}
\section{Acknowledgments}
\vspace{-2mm}

The computations have been carried out on the Brutus cluster at the ETH Z\"urich. 
This work is supported by the Swiss National Science Foundation under grant 200020-122117.
\vspace{-3mm}\\

\end{document}

%% file: newcommand.tex
\newcommand{\be}{\begin{equation}}
\newcommand{\ee}{\end{equation}}
\newcommand{\ba}{\begin{eqnarray}}
\newcommand{\ea}{\end{eqnarray}}

\newcommand{\nn}{\nonumber \\}

\newcommand{\Nf}{N_{\rm f}}

\newcommand{\Nc}{N_{\rm c}}

\def\lsi{\raise0.3ex\hbox{$<$\kern-0.75em\raise-1.1ex\hbox{$\sim$}}}
\def\gsi{\raise0.3ex\hbox{$>$\kern-0.75em\raise-1.1ex\hbox{$\sim$}}}

\newcommand{\Id}{\mathbbm1}

\newcommand{\expval}[1]{\left\langle #1 \right\rangle}

\newcommand{\Ocal}{\mathcal{O}}    
    
\newcommand{\Ord}[1]{\Ocal(#1)}

\newcounter{arabiclistc}

\def\sqr#1#2#3{{\vcenter{\hrule height.#2pt
      \hbox{\vrule width.#2pt height#1pt \kern#3pt
         \vrule width.#2pt}
      \hrule height.#2pt}}}